\documentclass[fleqn,10pt]{wlscirep}
\title{Concurrent enhancement of percolation and synchronization in adaptive networks}

\author[1,2]{Young-Ho Eom}
\author[3,4]{Stefano Boccaletti}
\author[1,5,6,7,*]{Guido Caldarelli}

\affil[1]{IMT School for Advanced Studies Lucca, Piazza San
Francesco 19, 55100 Lucca, Italy} \affil[2]{Departamento de
Matem\'{a}ticas, Universidad Carlos III de Madrid, 28911
Legan\'{e}s, Spain} \affil[3]{CNR-Istituto dei Sistemi Complessi,
Via Madonna del Piano, 10, 50019 Sesto Fiorentino, Italy}
\affil[4]{Italian Embassy in Israel, 25 Hamered Street, 68125 Tel
Aviv, Israel} \affil[5]{Istituto dei Sistemi Complessi (ISC), via
dei Taurini 19, 00185 Roma, Italy} \affil[6]{London Institute for
Mathematical Sciences, 35a South Street Mayfair, London, W1K 2XF,
UK} \affil[7]{Linkalab, Complex Systems Computational Laboratory,
Cagliari, Italy}

\affil[*]{guido.caldarelli@imtlucca.it}

\begin{abstract}
Co-evolutionary adaptive mechanisms are not only ubiquitous in
nature, but also beneficial for the functioning of a variety of
systems. We here consider an adaptive network of oscillators with
a stochastic, fitness-based, rule of connectivity, and show that
it self-organizes from fragmented and incoherent states to
connected and synchronized ones. The synchronization and
percolation are associated to abrupt transitions, and they are
concurrently (and significantly) enhanced as compared to the
non-adaptive case. Finally we provide evidence that only partial
adaptation is sufficient to determine these enhancements. Our
study, therefore, indicates that inclusion of simple adaptive
mechanisms can efficiently describe some emergent features of
networked systems' collective behaviors, and suggests also
self-organized ways to control synchronization and percolation in
natural and social systems.
\end{abstract}
\begin{document}

\flushbottom
\maketitle
%
%
\thispagestyle{empty}


\section*{Introduction}
Synchronization is possibly the paramount example of how
collective behaviors arise in complex systems, as it involves
emergence of collective organizations from microscopic
interactions of unitary constituents (such as neurons, heart
cells, power grids, or crickets~\cite{Pikovsky2003}). The
architecture of such interactions are formally well represented by
complex networks~\cite{Boccaletti2006,Caldarelli2007,Newman2010},
and underlying network structure of a system has, indeed, crucial
roles in synchronization~\cite{Arenas2008,Barrat2008}. For
instance, synchronization on small-world networks can be enhanced
compared to regular lattice thanks to the short average
distance~\cite{Barahona2002,Hong2004} while it could be more
difficult on scale-free networks compared to random homogeneous
networks due to increased concentration of load to highly
connected nodes~\cite{Nishikawa2003}. Also synchronization can
emerges more easily from weighted networks~\cite{Chavez2005} and
scale-free networks and Erd\"{o}s-Renyi networks follow different
paths to synchronization~\cite{GomezGardenes2007}.


The simplest approach to synchronization in networks is assuming a
static network structure. However, this approach does not
reproduce the behavior observed in real-world systems, where the
tendency observed is actually not static, rather dynamic. To cope
with this limitation, synchronization have been considered on
temporal or time-varying
networks~\cite{Lu2003,Belykh2004,Lee2012,Kohar2014}. For example,
systems of mobile oscillators have been introduced to consider
situations where interaction topology changes due to motion of the
oscillators~\cite{Frasca2008,Fujiwara2011,Uriu2013,Prignano2013}.
On the other hand, one can observe co-evolution of network
structure and network dynamics in many natural and social systems.
To take into account these co-evolutionary adaptive mechanisms,
various {\it adaptive} network models were
introduced~\cite{Gross2009}, where structure and the dynamics
co-evolve in time\cite{Gross2008,Garlaschelli2007}, and states of
the nodes shape the structure of their interaction, cooperatively
and simultaneously. Synchronization on adaptive networks has been
shown interesting phenomena~\cite{Aoki2009,Aoki2011}. Moreover
adaptive mechanisms are not only realistic, but they can also
enhance and stabilize collective
processes~\cite{Zhou2006,QRen2007,Sorrentino2008,Sorrentino2009},
change the order of synchronization~\cite{XZhang2015}, or enable
the emergence of meso-scale structures and scale-free
properties~\cite{Gutierrez2011,Assenza2011}.

Current studies on synchronization are, so far, focused on
completely percolated networks, i.e., in a situation where all
interacting oscillators belong to a single giant connected
component. However, real-world systems often show, even
temporarily, sparser and non-connected structures, as links
between units might well be not {\it continuously}
active~\cite{Holme2012,Gautreau2009}. In such non-connected
configurations (where not all nodes belong to a single connected
component), achieving global functions (e.g., synchronization) may
be hampered by the absence of stable interactions between the
units.

In this paper, we consider an adaptive network of oscillators,
where every unit (i.e., oscillator) selects its neighborhood on
the basis of a homophily principle~\cite{McPherson2001}.
Specifically, each oscillator is meant establishing connections
with the others that share a similar phase, in analogy to what
observed in social and natural systems~\cite{McPherson2001}. It is
worth noticing that such a {\it similarity} might be
time-dependent, as distinct oscillators adjust their phases but
also (and simultaneously) update the network structure following
homophily principles. We will show that our framework
qualitatively and quantitatively differs from non-adaptive
networks, in that synchronization and percolation transitions come
out to be substantially enhanced.

\section*{The adaptive network model}
We start by considering a network of $N$ (Kuramoto-type) phase
oscillators~\cite{Kuramoto1984,Strogatz2000}, whose time evolution
is ruled by:
\begin{equation}\label{eq:kuramoto}
\frac{d\theta_i}{dt} = \omega_i + \lambda \sum_{j=1}^N
a_{ij}\sin(\theta_j - \theta_i)
\end{equation}
where  $\omega_i$ ($\theta_i$) is the natural frequency (the
instantaneous phase) of oscillator $i$ drawn from a uniform
distribution in the range $[-1,1]$, $\lambda$ is the coupling
strength, and $\{a_{ij}\}$ are the elements of the network's
adjacency matrix.

The structure of connections is given by the so-called {\it
fitness} or {\it hidden variable} network
model~\cite{Caldarelli2002,Garlaschelli2004}, which is a
generalized Erd\"{o}s-Reyni (ER) model. The distinctive character
of such a model is that the topology is fully shaped by the
fitness of the nodes (herein associated to the oscillators'
phases) while the topology is given by a constant probability in
the ER model. Accordingly the connection probability between two
node $i$ and $j$ at time $t$ is determined by a given function
$f(\theta_i,\theta_j)$. While the form of function $f$ can be, in
general, arbitrary, we here consider it to follow a homophily
principle, through which oscillators with more similar phases are
more likely to be connected. For the sake of exemplification, we
then define the function $f$ as follows:

\begin{equation}\label{eq:topology}
f(\theta_i,\theta_j) = \frac{z(1+\cos(\theta_i-\theta_j))}{N}
\end{equation}

where $z$ is a positive parameter, $f(\theta_i,\theta_j)=2z/N$ if
$\theta_i = \theta_j$ and $f(\theta_i,\theta_j)=0$ if
$|\theta_i-\theta_j|=\pi$. If two oscillators feature close enough
phases (i.e., $|\theta_i-\theta_j| \sim 0$), they are then more
likely to establish a link, with probability $2z/N$. The
expectation is therefore that higher $z$ values would lead to more
connected network structure, while higher $\lambda$ values would
result into more coherent dynamical state. We assume that at each
time step the phases of oscillators are updated by
Eq.~\ref{eq:kuramoto} and at the same time step, with a coupling
probability $P$, the network topology is shaped by
Eq.~\ref{eq:topology}. In this study, without specific indication,
we consider the case of $P=1.0$. For comparison we show the
results with $P=0.5$ and $P=0.2$, which are very similar with the
case of $P=1.0$, in the Supplementary Information.

\section*{Results}

In our simulations, performed with a 4th order Runge-Kutta method
and a time-step $\Delta t=0.02$ (See the Supplementary Information
for the case of $\Delta t=0.05$ and $\Delta t=0.1$ for
comparison), we consider a network size $N=300$ (See the
Supplementary Information for cases of $N=150$ and $N=600$). We
assign initial conditions for the oscillators' phases from a
uniformly distributed distribution in the range $[-\pi,\pi]$,
while the initial network structure is taken to be that extracted
from Eq.~(\ref{eq:topology}) with the given initial phases. At
each time step of the integration, oscillators' phases evolve by
Eq.~(\ref{eq:kuramoto}), and (simultaneously) network structure is
reshaped by Eq.~(\ref{eq:topology}). To compare with, the
non-adaptive evolution is also simulated, where the structure of
the network is determined by Eq.~(\ref{eq:topology}) {\it only
initially}.

The degree of synchronization can be monitored by the
synchronization order parameter:
\begin{equation}
r(t)e^{i\Psi(t)} = \frac{1}{N}\sum_{j=1}^N e^{i\theta_j(t)},
\end{equation}
whose modulus ($r(t) \in [0,1]$) measures actually the system's
phase coherence ($r=1$ for the fully synchronized regime, $r\sim0$
for the incoherent state). $\Psi(t)$ is instead the average phase
of the system. For percolation, we consider the relative size of
the largest connected component $s(t)$  as the order parameter.
For each parameter $r(t)$ and $s(t)$, we furthermore define $R$
and $S$ as the respective steady state values, i.e. the values
obtained by averaging over 500 steps, and after 3,000 transient
steps.

\begin{figure}[ht]
\centering
\includegraphics[width=0.7\linewidth]{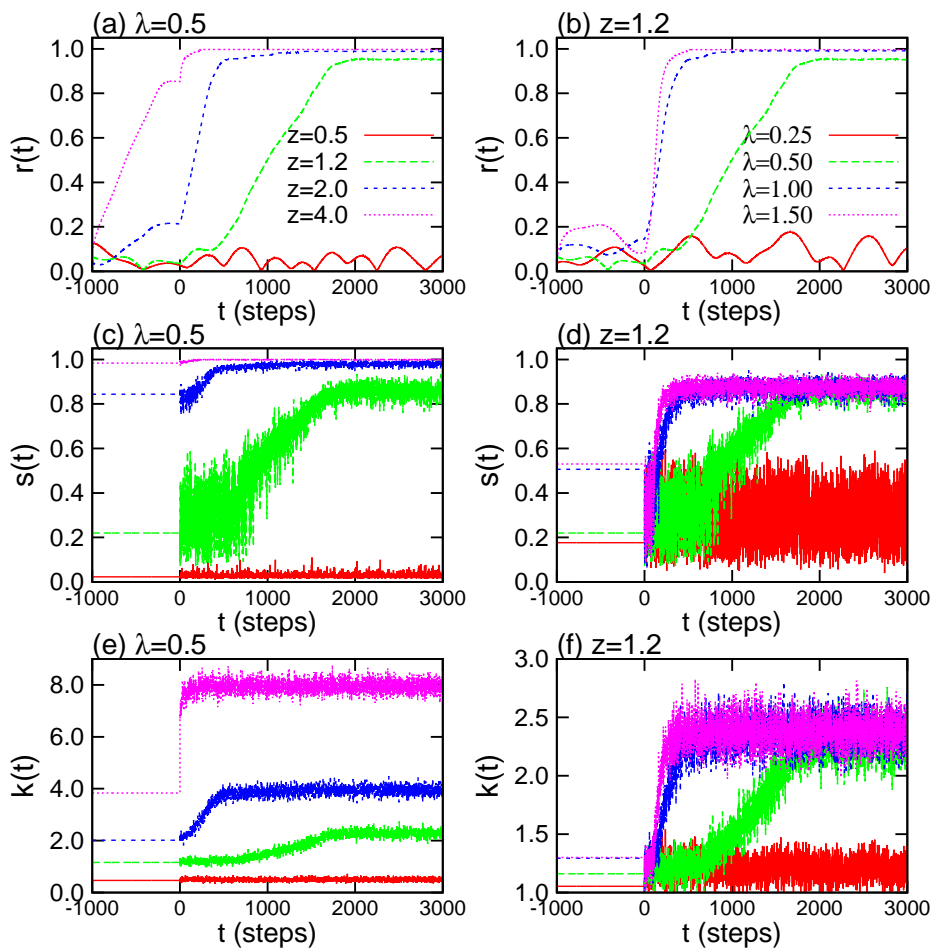}
\caption{Time evolution of $r(t)$ (a-b), $s(t)$ (c-d) and of the
network's average degree $k(t)$ (e-f).  (a, c, and e)
$\lambda=0.5$; (b, d, and f) $z=1.2$. Color codes in the legends
of (a) and (b).} \label{fig:1}
\end{figure}

Figure~\ref{fig:1} reports the time evolution of $r(t)$ and
$s(t)$, at different values of the control parameters $z$ and
$\lambda$. When $t<0$, the time evolution of the order parameters
is determined by the fixed network structure constructed by
Eq.~\ref{eq:topology} with the initial phases (i.e., non-adaptive
networks), whereas the network structure (starting from $t=0$) is
updated by Eq.~\ref{eq:topology} at every time step. In
Fig.~\ref{fig:1}(a) and (c), $r(t)$ and $s(t)$ are plotted at
$\lambda=0.5$ and varying $z$, respectively while
Fig.~\ref{fig:1}(b) and (d) reports $r(t)$ and $s(t)$ (at fixed
$z=1.2$) by varying $\lambda$. A clear enhancement of
synchronization and percolation is simultaneously observed for
most values of $\lambda$ and $z$ (except when $z=0.5$ and
$\lambda=0.5$, or when $z=1.2$ and $\lambda=0.25$). The evolution
of the network's average degree $k(t)$ [Figs.~\ref{fig:1}(e) and
(f)] reveals that adaptation leads actually to an increase of the
average degree, which may explain the concurrent enhancement of
percolation and synchronization in the adaptive network.

\begin{figure}[ht]
\centerline{\includegraphics[width=0.7\linewidth]{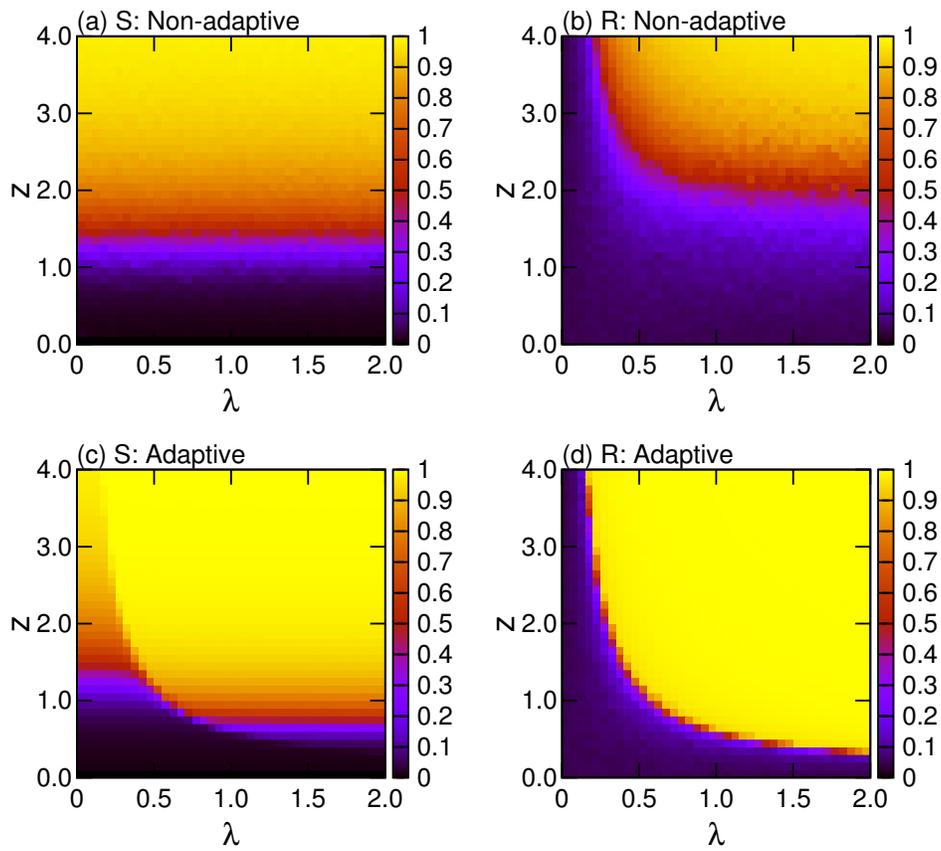}}
\caption{Phase diagrams of the non adaptive (a,b) and
adaptive(c,d) models. Panels refer to the percolation indicator
$S$ (a,c) and the synchronization indicator $R$ (b,d). For each
$z$ and $\lambda$, data refer to ensemble averages over $40$
different realizations. } \label{fig:2}
\end{figure}

Figure~\ref{fig:2} accounts for $S$ and $R$ in the parameter space
($\lambda$, $z$). The percolation transition in the non-adaptive
network only depends on $z$ [as shown in Fig.~\ref{fig:2}(a)]. We
observe existence of typical percolation transitions within the
subcritical regime ($S\sim 0.0$) of $z<1.0$, the critical regime
of $z \sim 1.0$, and the supercritical regime ($0.0<S<1.0$) of
$1.0<z<3.0$, and also the connected regime ($S\sim 1.0$) is
observed for $z>3.0$. As shown in Fig.~\ref{fig:2}(b),
synchronization in the non-adaptive case depends on the specific
percolation state the network is attaining. Fully incoherent
states ($R <0.05$) are observed in sub-critical and critical
regime ($z<1.0$) regardless of $\lambda$. Partial synchronization
($0.1<R<0.9$) is observed, instead, in supercritical regimes, and
highly synchronized states emerge only in the connected regime
($z>3.0$).

On the other hand, significant enhancement of percolation and
synchronization is evident in Figs.~\ref{fig:2}(c) and (d). In
particular, the enhancement is substantial in the region of
$z<3.0$ corresponding to the non-connected regimes in the
non-adaptive network. In particular, the percolation indicator $S$
depends not only on $z$, but also on $\lambda$, and  (when
$\lambda$ increases) the giant connected component emerges even
for smaller values of $z$.

Furthermore, synchronization is actually boosted in the adaptive
network [Fig.~\ref{fig:2}(d))]. Similarly to percolation, the
enhancement is here predominant in low connection ability regions
($z<3.0$). Interestingly enough, also some not-fully connected
regions ($S<1.0$) still can display highly coherent states ($R
\sim 1$). The conclusions that can be drawn from our results is
that the adaptive mechanism actually creates a positive feedback
loop between network's structure and dynamics, thence supporting
the ubiquity of synchronized and connected components in complex
systems under limited resources for interactions.

\begin{figure}[ht]
\centerline{\includegraphics[width=0.7\linewidth]{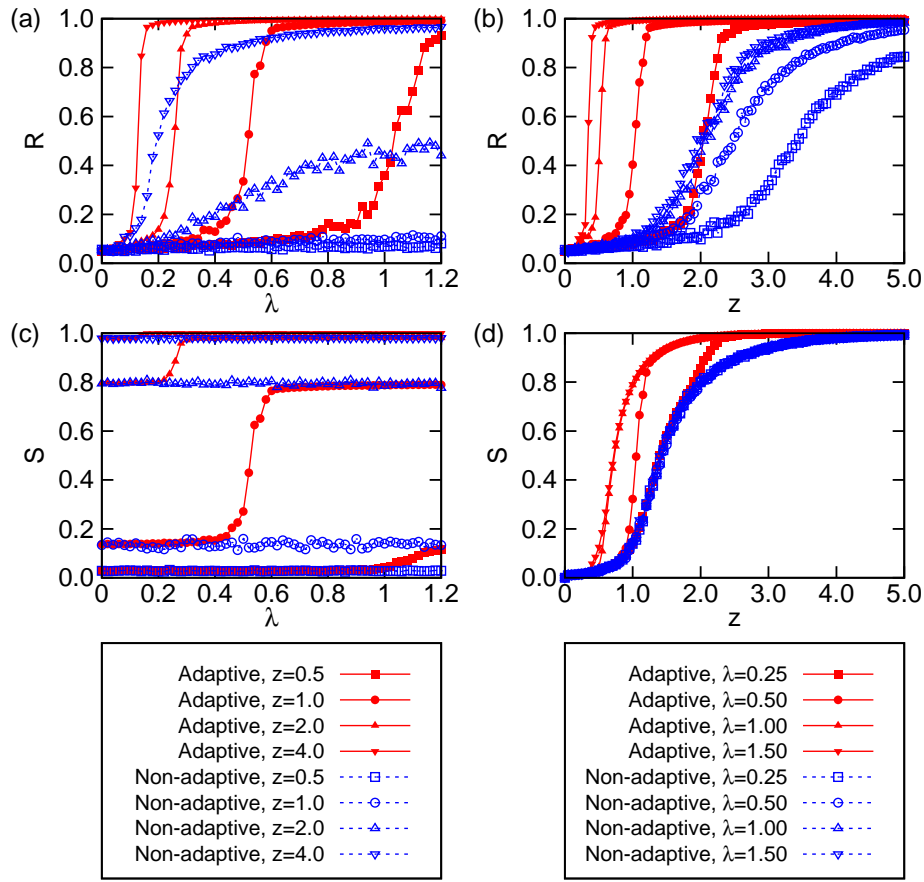}}
\caption{$R$ (a,b) and $S$ (c,d) for adaptive and non-adaptive
networks. (a) $R$ {\it vs.} $\lambda$ at different $z$ values; (b)
$R$ {\it vs.} $z$ at different $\lambda$ values; (c) $S$ {\it vs.}
$\lambda$ at different $z$ values; (d) $S$ {\it vs.} $z$ at
different $\lambda$ values. Legends (in the bottom panels) have to
be referred to for the understanding of the used parameters'
values. Data refer to ensemble averages over $40$ realizations.}
\label{fig:3}
\end{figure}

The adaptive mechanisms here considered not only enhance
synchronization and percolation, but also make both transitions
more abrupt. In other words both transitions in the adaptive
networks are more sensitive to the coupling strength $\lambda$ and
to the connectivity parameter $z$ than the transitions in the
non-adaptive networks. Note that, in this sense, here we do not
consider the observed transitions as so-called explosive
synchronization~\cite{GomezGardenes2011} or
percolation~\cite{Achlioptas2009}. In Figure~\ref{fig:3} we report
$R$ [panels (a) and (b)] and $S$ [panels (c) and (d)] as a
function of $\lambda$ at fixed $z$ , as well as varying $z$ at
fixed $\lambda$.  For non-adaptive networks, the passage from
incoherent to coherent states (and that from fragmented to
percolated structures) features typical traits of second-order
transitions, while adaptive networks displays abrupt patterns. The
case of percolation transition shows, actually, more interesting
patterns. When $z$ is fixed, $S$ in the non-adaptive network does
not depend on $\lambda$ [as shown in Fig.~\ref{fig:2}(a) and
Fig.~\ref{fig:3}(c)]. However, $S$ in the adaptive case shows a
clear percolation transition with growing $\lambda$ when $z<4.0$
[see the red lines with filled symbols in Fig.~\ref{fig:3}(c)].
Interestingly, there is no difference in the behavior of $S$
(before the transition) between the adaptive and non-adaptive
case. Only above certain values of $\lambda$, the percolation
transition assumes a characteristic "first-order-type nature" [as
seen in Fig.~\ref{fig:3}(d)]. It is notable that, although the
interplay between network evolution and dynamics happens here
simultaneously, the transition to synchrony seems to occur at
lower $z$ or $\lambda$ values, actually, than the percolation
transition.

\begin{figure*}[ht]
\begin{center}
\includegraphics[width=0.96\linewidth,angle=0]{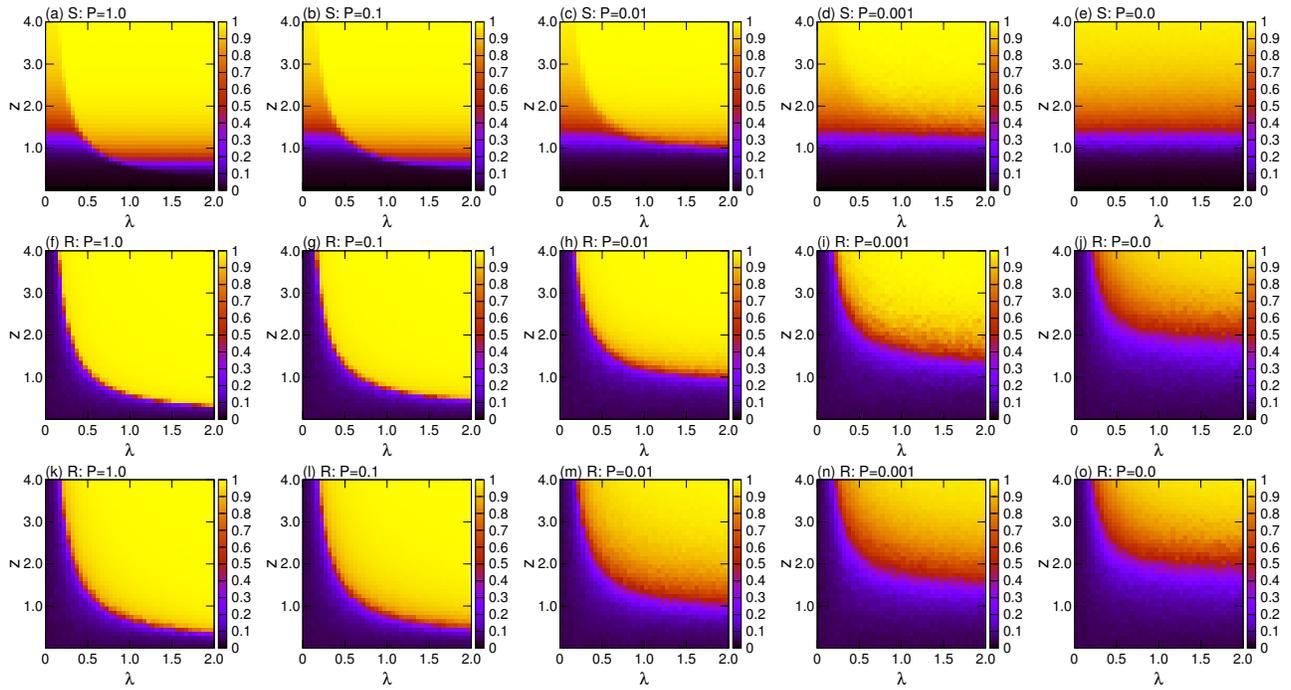}
\caption{ $S$ (top row) and $R$ (middle row) in the parameter
space ($z,\lambda$) for the adaptive network with different
coupling probability $P$. Bottom row reports, instead, $R$ (in the
same parameter space) for blinking networks with different
coupling probability $P$. Once again, data refer to ensemble
average over $40$ different realizations for each $z$ and
$\lambda$. } \label{fig:4}
\end{center}
\end{figure*}

While the effect of the interplay between topological and
dynamical evolution of nodes appears to be clear, it is of the
highest importance orienting the study to the inspection of  the
timescales at which the two phenomena appear. In particular, if
updating network structure costs more than updating states of
oscillators, it is necessary to check whether adaptive mechanisms
should be applied at every time step or, instead, just few
applications of them are actually sufficient to determine the
observed enhancements. The issue is here addressed by introducing
a coupling probability $P$ between dynamics of oscillators and
structural evolution, namely by updating the network structure
[via Eq.~(\ref{eq:topology})] with probability $P$ at each time
step. The limit $P=0$ recovers a non-adaptive network model, while
$P=1.0$ corresponds to a totally adaptive case. In
Fig.~\ref{fig:4} we report $S$ (top row) and $R$ (middle row) from
the cases of $P=1$, $0.1$, $0.01$, $0.001$ and $0$. Remarkably,
one observes that both transitions (to percolation and synchrony)
are significantly enhanced along all the finite range of $P$,
including $P=0.001$. This fact has significant implications, in
the sense that one can actually intervene on the collective
behaviors of a given system, only with a few applications of our
proposed adaptive mechanism.

It was recently reported that blinking networks (i.e. topologies
of interactions which change over timescale much faster than that
of the network units' dynamics), can actually enhance
synchronization~\cite{Hasler2013,Hasler2013a}. As our adaptive
model also can have such a 'blinking' nature (when $P \sim 1.0$),
it is therefore mandatory to comparatively investigate on how much
the observed enhancement in synchronization has a route within the
yet known blinking effects. To this purpose, we consider a
blinking network of oscillators (which is exactly the same as the
considered adaptive network) with a topology updated by a random
probability $Q$, and which gives the same number of links at the
initial step given by Eq.~\ref{eq:topology}. Note that whether
updating topology or not at each time step depends on the coupling
probability $P$ in both of the adaptive network and the blinking
network while the connections between the oscillators are given by
Eq.~\ref{eq:topology} in the adaptive network but by the random
probability $Q$ in the blinking network. The bottom panels of
Fig.~\ref{fig:4} reports the values of $R$ for such a latter,
blinking, network as function of $\lambda$ and $z$, with varying
$P$. When $P=1.0$, one notices that the blinking effect is,
indeed, quite strong. However, the effect vanishes rapidly with
decreasing $P$. This indicates that our adaptive mechanism may
enhance synchronization {\it only partially} due to blinking
effects, whereas significant other contributions exist. It is also
noticeable that no enhancement in percolation exists at all in the
blinking framework, due punctually to the lack of feedback between
dynamics of oscillators and topological evolution.




\section*{Discussion}
In conclusion, complex networks need to stay in connected and
synchronized states, in order to perform integrated and coherent
functions. However, when the units have only limited ability to
connect to each other, it is of paramount importance understanding
how the networks self-organize from fragmented and incoherent
states to connected and synchronized states. We have considered an
adaptive model, where connections between nodes are ruled by a
positive feedback loop connecting structural evolution (driven by
a fitness model) and nodal dynamics (driven by the Kuramoto
model). We actually gave evidence that such an adaptive framework
enhances substantially synchronization and percolation, while
non-adaptive counterparts fail to reach synchronization and
percolation in the non-connected regime. This indicates that
co-evolutionary adaptive networks are not only more realistic
descriptions of complex systems, but also they are beneficial for
the correct and robust functioning of complex systems.

The observed enhancement of synchronization and percolation shed
actually light on how one can control such two processes in a
spontaneous, or self-organized, way~\cite{Garlaschelli2007}. In
particular, as shown in our Fig.~\ref{fig:4}, the needed coupling
has not to be very strong, thus suggesting that the control of
unwanted events emerging through synchronization (such as
epileptic seizure or market crashes) could be easily achieved by
just (properly) coupling or decoupling network's structure
evolution and dynamics. In this sense, our findings suggest
efficient control methods to maintain an integrated functioning of
natural and social systems.


\bibliography{ScientificCitationRef}

\section*{Acknowledgements (not compulsory)}
Y.-H. E. and G.C. acknowledge FET Project MULTIPLEX (nr. 317532),
FET Project SIMPOL (nr. 610704) and FET Project DOLFINS (nr.
640772). Y.-H. E. would like to thank the Universidad Carlos III
de Madrid, the European Union's Seventh Framework Programme for
research, technological development and demonstration under grant
agreement (n\textsuperscript{\b{o}} 600371), el Ministerio de
Econom\'{i}a y Competitividad (COFUND2014-51509) and Banco
Santander

\section*{Author contributions statement}
Y.-H.E, S.B, and G.C. conceived the experiment(s), Y.-H.E
conducted the experiment(s), Y.-H.E, S.B, and G.C. analysed the
results, wrote the manuscript, and reviewed the manuscript.

\section*{Additional information}
\textbf{Competing financial interests}: The authors declare no
competing financial interests.


\end{document}